\documentclass{article}
\usepackage{spconf,amsmath,graphicx,hyperref}
\usepackage{tabularx}
\usepackage{subcaption}
\usepackage{fontspec}
\usepackage{float}

\newfontfamily\devanagarifont[Script=Devanagari]{NotoSerifDevanagari-VariableFont_wdth,wght.ttf}

\title{Neural Multi-Speaker Voice Cloning for Nepali in Low-Resource Settings}
\name{%
Aayush M. Shrestha$^{\ast}$, 
Aditya Bajracharya$^{\ast}$, 
Projan Shakya$^{\ast}$, 
Dinesh B. Kshatri%
\thanks{$^\ast$These authors contributed equally to this work. \\}%
}

\address{IOE, Thapathali Campus, Kathmandu, Bagmati Province, Nepal \\
\{aayush.077bct003, aditya.077bct006, projan.077bct037\}@tcioe.edu.np, dinesh@ioe.edu.np}

\begin{document}
%
\maketitle
\begin{abstract}
This research presents a few-shot voice cloning system for Nepali speakers, designed to synthesize speech in a specific speaker’s voice from Devanagari text using minimal data. Voice cloning in Nepali remains largely unexplored due to its low-resource nature. To address this, we constructed separate datasets: untranscribed audio for training a speaker encoder and paired text-audio data for training a Tacotron2-based synthesizer. The speaker encoder, optimized with Generative End2End (GE2E) loss, generates embeddings that capture the speaker’s vocal identity, validated through Uniform Manifold Approximation and Projection (UMAP) for Dimension Reduction visualizations. These embeddings are fused with Tacotron2’s text embeddings to produce mel-spectrograms, which are then converted into audio using a WaveRNN vocoder. Audio data were collected from various sources, including self-recordings, and underwent thorough preprocessing for quality and alignment. Training was performed using mel and gate loss functions under multiple hyperparameter settings. The system effectively clones speaker characteristics even for unseen voices, demonstrating the feasibility of few-shot voice cloning for the Nepali language and establishing a foundation for personalized speech synthesis in low-resource scenarios.
\end{abstract}
\section{Introduction}
\label{sec:intro}

Voice cloning has advanced significantly with the rise of deep learning, enabling the synthesis of natural-sounding human speech directly from text. While substantial progress has been achieved for high-resource languages like English, low-resource languages such as Nepali have received comparatively little attention. Despite the growing adoption of voice cloning in areas such as entertainment, accessibility, and personalized voice assistants, most existing solutions remain confined to English, highlighting a significant research gap. Nepali, a linguistically rich and culturally diverse language, is particularly underrepresented in speech synthesis research. This lack of technological development is especially striking given the variety of dialects, regional accents, and culturally significant voices found across Nepal.

This paper addresses this gap by developing a neural voice cloning system tailored for the Nepali language. Our objective is to enable multi-speaker speech synthesis using a dataset of Nepali voices, thereby contributing to linguistic diversity within the field of speech technology.  Furthermore, the development of such systems could enhance accessibility for individuals with speech impairments and promote the preservation and revitalization of Nepal’s vocal heritage.

The main contributions of this work are as follows:
\begin{enumerate}
    \item We collected and curated a clean, structured Nepali speech dataset that is ready for direct use in Text-to-Speech (TTS) and voice cloning research.
    \item We trained a multi-speaker generative model capable of cloning voices in the Nepali language, demonstrating the feasibility and effectiveness of neural speech synthesis for a low-resource linguistic context.
\end{enumerate}

\section{Related Works}
\label{sec:related}

Voice cloning has been extensively studied in English, with pioneering work by ~\cite{chen2018voicecloning} proposing a multi-speaker generative model capable of few-shot adaptation. Their approach utilizes a speaker encoder trained to generalize across speakers, enabling synthesis from a few untranscribed audio samples. Two strategies, Speaker Adaptation and Speaker Encoding, were introduced to balance naturalness and efficiency. ~\cite{jia2018transfer} further advanced this with the SV2TTS framework, consisting of a speaker verification network trained with GE2E loss, a Tacotron 2-based synthesizer, and a WaveNet vocoder. This model demonstrated robust multi-speaker text-to-speech synthesis, including voices of unseen speakers.

Research efforts have also extended voice cloning to other languages. ~\cite{liu2021metavoice} introduced \textit{Meta-Voice}, a fast few-shot style transfer TTS for expressive voice cloning in Chinese, leveraging meta-learning to adapt with fewer than five samples. ~\cite{zhang2022multimodalvc} proposed a multi-modal few-shot voice cloning system for Mandarin that integrates unsupervised speech representations to improve synthesis quality. ~\cite{zhao2022nnspeech} presented \textit{nnSpeech}, a zero-shot multi-speaker TTS system using a conditional variational autoencoder, validated on English and Mandarin datasets.

In the open-source community, CorentinJ’s Real-Time Voice Cloning toolkit~\cite{corentinj2019realtime} implements the SV2TTS pipeline, enabling real-time voice cloning from as little as five seconds of audio. Our work adapts this architecture for the Nepali language, an underrepresented language in voice cloning research, by fine-tuning the model on a limited Nepali dataset.

While multilingual and zero-shot voice cloning models have been emerging, most research to date has focused on high-resource languages. Our study addresses this gap by exploring voice cloning for Nepali, leveraging proven architectures and adapting them to a low-resource setting.

We found several Nepali text-to-speech (TTS) projects that share some functional similarities with our system. One such project is Shruti: A Nepali Book Reader \cite{khadka2023tts}, which reads text from books provided in PDF format by first extracting the text using Optical Character Recognition (OCR).  

Another project in this domain is Aawaj: Augmentative Communication Support for the Vocally Impaired using Nepali Text-to-Speech \cite{aawaj}. This system also focuses on Nepali TTS, but lacks the capability of multi-speaker modeling or voice cloning.  

In addition to TTS efforts, a study by \cite{karki2024advancingvoicecloningnepali} explored Nepali voice cloning using WaveNet as the vocoder. While this work represents an initial step toward voice cloning in Nepali, it was limited to training on pre-existing data from the OpenSLR corpus.  

In comparison, our research moves beyond conventional TTS and prior voice cloning attempts by incorporating multi-speaker voice cloning, enabling the synthesis of speech in voices not limited to a single pre-trained speaker. This distinction highlights the novelty of our approach with respect to previous Nepali speech synthesis efforts. 

\section{Dataset Description}
\label{sec:dataset_desc}

We need datasets for two components: speaker encoder and synthesizer. 
For speaker encoder, we focus on variation and quantity i.e. to train speaker encoder, we require Nepali audio from many speakers. Noise is tolerable up to a certain degree.  We had to collect audio data for different speakers with different variations to ensure our program could clone different variations of linguistic features.
For synthesizer, we focus on quality i.e. to train synthesizer, we require Nepali dataset consisting of a pair of audios along with the equivalent transcript. The audio must have minimum to no noise and the transcript must be accurate.

\subsection*{Dataset for Speaker Encoder}
To train our speaker encoder, we utilized a combination of publicly available and self-curated datasets to ensure a diverse and representative coverage of speaker characteristics across varying domains and acoustic conditions.

\noindent\textbf{OpenSLR Datasets}
\begin{itemize}
  \setlength\itemsep{0em} 
  \item SLR43 \cite{openslr_slr43}
  \item SLR54 \cite{openslr_slr54}
  \item SLR143 \cite{khadka2023tts}
\end{itemize}

\noindent\textbf{Self-Collected Datasets} \\
Sources include:
\begin{itemize}
  \setlength\itemsep{0em}
  \item Audiobooks
  \item Self-recorded speech
  \item YouTube content such as interviews and podcasts
\end{itemize}

\begin{figure}[htbp]
    \centering
    \includegraphics[width=\columnwidth]{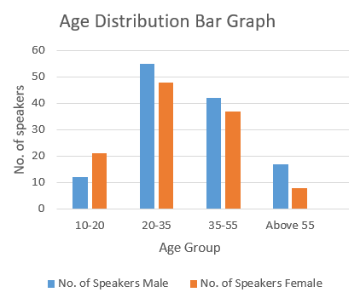}
    \caption{Age distribution of male and female speakers across different age groups for our self-collected dataset for speaker encoder}
    \label{fig:age-distri}
\end{figure}

To minimize variability across different age groups, we attempted to collect speech samples from a wide range of participants. However, as illustrated in Figure~\ref{fig:age-distri}, the dataset is skewed toward individuals aged 20–35 years, owing to their greater availability. In contrast, collecting data from speakers younger than 20 years and older than 55 years proved considerably more challenging, resulting in under representation of these age groups. 

\begin{figure}[!t]
    \centering
    \includegraphics[width=\columnwidth]{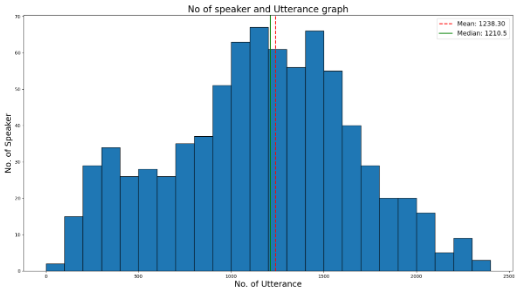}
    \caption{Average number of utterances per speaker.}
    \label{fig:hist}
\end{figure}

Figure~\ref{fig:hist} shows the distribution of utterances per speaker, with most speakers contributing between 800 and 1600 utterances. The mean and median are 1238.3 and 1210.5, respectively, marked by the red and green lines. On average, each speaker contributed about 1238 utterances, obtained by segmenting audio into 1.6-second chunks with 50\% overlap. The distribution is relatively balanced across speakers, though a few contributed notably fewer or more utterances, which helps mitigate speaker-specific bias during model training.

The finalized speaker encoder dataset (Table~\ref{tab:speaker_encoder_summary}) comprises recordings from 833 distinct speakers, including 481 male and 352 female participants. It spans a cumulative duration of 235 hours, distributed across 6,466 utterances. The duration of individual speaker contributions ranges from 49 seconds to 58 minutes and 51 seconds, ensuring a wide distribution of speaking styles and voice characteristics. This diversity is critical for training speaker embeddings capable of capturing inter-speaker variability across a broad spectrum of vocal features.

\begin{table}[t]
\centering
\caption{Summary Statistics of Final Speaker Encoder Dataset}
\label{tab:speaker_encoder_summary}
\begin{tabularx}{\columnwidth}{l >{\centering\arraybackslash}X}
\hline
\textbf{Metric} & \textbf{Value} \\
\hline
Total Speakers                  & 833 \\
Male Speakers                   & 481 \\
Female Speakers                 & 352 \\
Total Duration                  & 235 hours \\
Longest Audio (Single Speaker)  & 58 minutes 51 seconds \\
Shortest Audio (Single Speaker) & 49 seconds \\
Total Number of Utterances      & 6,466 \\
\hline
\end{tabularx}
\end{table}

\subsection*{Dataset for Synthesizer}

To train our speech synthesizer, we employed a combination of publicly available datasets alongside self-collected recordings to capture a broad range of speaker variability and recording conditions. This approach ensures the model benefits from both high-quality, controlled data and more diverse, real-world audio.

\vspace{1em}
\noindent\textbf{OpenSLR Datasets}
\begin{itemize}
  \setlength\itemsep{0em}
  \item SLR43~\cite{openslr_slr43}
  \item SLR143~\cite{khadka2023tts}
\end{itemize}

\vspace{1em}
\noindent\textbf{Self-Collected Datasets}
\begin{itemize}
  \setlength\itemsep{0em}
  \item Audiobooks
  \item Self-recorded speech
\end{itemize}

\begin{figure}[!t]
    \centering
    \includegraphics[width=\columnwidth]{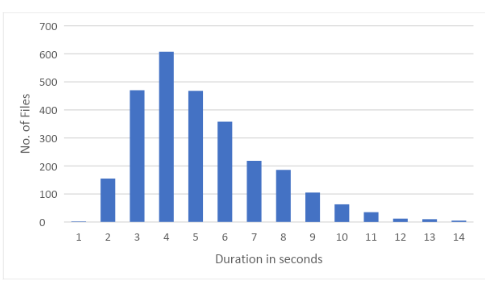}
    \caption{Audio duration of the collected dataset and the number of audio files for Synthesizer data}
    \label{fig:audio-len}
\end{figure}

Figure~\ref{fig:audio-len} presents the distribution of audio file durations in the dataset. The x-axis represents the duration of audio files in seconds, ranging from 1 to 14 seconds, while the y-axis shows the corresponding number of files. The histogram indicates that most audio files are concentrated between 3 and 6 seconds, with the highest frequency observed at 4 seconds. The number of files decreases gradually for durations longer than 6 seconds, suggesting that shorter audio clips are more common in the collected dataset.

The synthesizer dataset (detailed in Table~\ref{tab:synthesizer_dataset_summary}) includes 6,046 audio-text pairs, totaling 64,232 words and 407,371 characters over 8.67 hours of speech. Clip durations vary between 1.404 and 14.582 seconds, with a mean length of 5.162 seconds. On average, each clip contains 10.63 words, contributing to a total of 79,375 distinct lexical entries. This dataset is designed to support high-fidelity speech synthesis by providing a phonetically and prosodically rich corpus suitable for modeling nuanced vocal expressions.

\vspace{1em}
\begin{table}[!t]
\centering
\caption{Final Summary of Synthesizer Dataset}
\label{tab:synthesizer_dataset_summary}
\begin{tabular}{l c}
\hline
\textbf{Metric} & \textbf{Value} \\
\hline
Total Clips              & 6,046 \\
Total Words              & 64,232 \\
Total Characters         & 407,371 \\
Total Duration           & 8.67 hours \\
Mean Clip Duration       & 5.162 seconds \\
Minimum Clip Duration    & 1.404 seconds \\
Maximum Clip Duration    & 14.582 seconds \\
Mean Words per Clip      & 10.63 \\
Distinct Words           & 79,375 \\
\hline
\end{tabular}
\end{table}

\section{Dataset Preprocessing}
\label{sec:preprocess}

\subsection*{Audio Processing}
Audio preprocessing was conducted using Audacity and custom Python scripts to standardize and enhance the audio data. Initially, all files were converted to the common .wav format from various original formats such as .flac, .m4a, and .mp3. To maintain uniformity and prevent information loss, audio clips were resampled to 16 kHz for the speaker encoder and 22.05 kHz for the synthesizer. Stereo recordings were converted to mono to simplify processing. Silence segments longer than 0.5 seconds were truncated to 0.1 seconds to reduce unnecessary pauses. Since noise is particularly detrimental for the synthesizer, breathing sounds and noisy signals were manually removed using Audacity. Normalization was applied across all audio clips to ensure consistent amplitude and minimize volume variations. Clips exceeding 15 seconds were split into two separate segments. For the speaker encoder, audio clips were further segmented into 1.6-second chunks with 50 overlap to maximize data utilization.

\subsection*{Text Preprocessing}

Text preprocessing was carried out to impose consistency and linguistic regularity across the corpus. Instances of numerals, years, and dates expressed in numeric form were converted into their full textual realizations, thereby mitigating representational discrepancies and facilitating word-based rather than digit-based modeling. To address structural irregularities, complex or multi-clausal sentences were segmented into shorter, syntactically coherent units, which improves boundary detection and enhances downstream processing. In addition, where sentence-final markers were absent or inconsistently used, appropriate stop tokens were manually inserted to ensure explicit demarcation of sentence boundaries. Collectively, these procedures contributed to greater grammatical conformity and semantic clarity within the dataset. Representative examples of these preprocessing operations are provided in Table~\ref{tab:preprocessing_examples}, which illustrates numeral and date conversion, sentence segmentation, and manual stop token insertion in Nepali text.

\vspace{1em}
\begin{table*}[!t]
\renewcommand{\arraystretch}{1.5}
\small
\centering
\caption{Examples of Preprocessing Steps}
\label{tab:preprocessing_examples}
\begin{tabularx}{\textwidth}{>{\raggedright\arraybackslash}p{0.23\textwidth} >{\raggedright\arraybackslash}X >{\raggedright\arraybackslash}X}
\hline
\textbf{Preprocessing Type} & \textbf{Raw Text} & \textbf{Pre-processed Text} \\
\hline

Numeral and Date Conversion &
{\devanagarifont भर्खर ७ प्याग हुँदैछ, १५ प्याग खाएपछि मात्र मलाई थोरै थोरै लाग्छ ।} &
{\devanagarifont भर्खर सात प्याग हुँदैछ, पन्ध्र प्याग खाएपछि मात्र मलाई थोरै थोरै लाग्छ ।} \\

Sentence Segmentation &
{\devanagarifont बरु अब त अझ बढी लजालु भएकी थिएँ; साथीहरु पनि कम थिए ।} &
{\devanagarifont बरु अब त अझ बढी लजालु भएकी थिएँ। साथीहरु पनि कम थिए ।} \\

Manual Stop Token Addition &
{\devanagarifont त्यो चिच्याहट सुनेर क्यालीगुला आनन्दित हुन्थ्यो} &
{\devanagarifont त्यो चिच्याहट सुनेर क्यालीगुला आनन्दित हुन्थ्यो ।} \\

\hline
\end{tabularx}
\end{table*}

\section{Methodology}
\label{sec:method}

\begin{figure}[!t]
\centering
\includegraphics[width=\columnwidth]{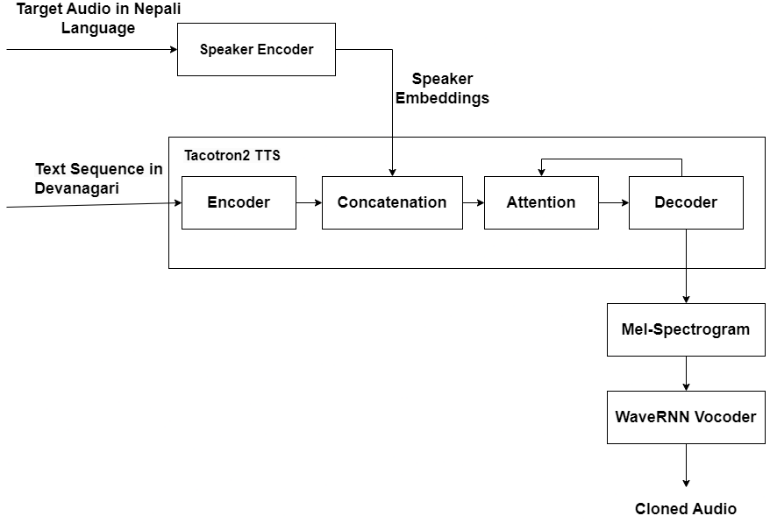}
\caption{Architecture of the Nepali voice cloning system combining a Speaker Encoder, Tacotron2-based TTS synthesizer, and WaveRNN vocoder.}
\label{fig:architecture}
\end{figure}

The architecture depicted in Figure~\ref{fig:architecture} illustrates the overall workflow of our Nepali voice cloning system. It begins with the input of target audio spoken in the Nepali language, which is processed by the Speaker Encoder to extract speaker embeddings that capture the unique vocal characteristics of the speaker. Simultaneously, the system takes the corresponding text sequence in Devanagari script as input to the Tacotron2-based Text-to-Speech (TTS) model. Within the Tacotron2 framework, the text is first encoded by the Encoder module into a hidden representation. This encoded text representation is then concatenated with the speaker embeddings, effectively conditioning the synthesis on the speaker’s identity. The combined features pass through an Attention mechanism that aligns and focuses on relevant parts of the text encoding during decoding. The Decoder generates a mel-spectrogram from this attended representation, representing the intermediate audio feature in the time-frequency domain. Finally, the mel-spectrogram is transformed into time-domain cloned audio by the WaveRNN vocoder, which synthesizes natural-sounding speech consistent with the target speaker’s voice. This pipeline integrates speaker verification techniques with state-of-the-art neural TTS models to achieve voice cloning for Nepali with limited data.

\subsection{Speaker Encoder}

The speaker encoder extracts fixed-dimensional embeddings that uniquely characterize a speaker’s voice, following the d-vector architecture. It employs three stacked Long Short-Term Memory (LSTM) layers with 256 units each, followed by a fully connected layer that projects the features into a 256-dimensional embedding space.

\subsubsection*{Model Architecture and Parameters}

The speaker encoder consists of three LSTM layers with 256 units each, followed by a fully connected layer producing 256-dimensional speaker embeddings. Key hyperparameters used during training are shown in Table~\ref{tab:model_hyperparams} of Appendix \ref{sec:appendixA} section.

\subsection{Synthesizer}

The synthesizer is based on the Tacotron architecture \cite{wang2017tacotron} and generates mel spectrograms conditioned on both textual input and speaker identity. Speaker embeddings extracted from the encoder are concatenated with the corresponding text embeddings to form a joint representation, which guides the spectrogram generation process.

\subsubsection*{Training Dataset and Configuration}

For fine-tuning, the synthesizer was trained on a curated dataset comprising a blend of Nepali TTS corpora and supplementary self-recorded audio. The dataset spans multiple speaker identities and acoustic environments, with recordings sampled primarily at 22.05 kHz and 48 kHz, as summarized in Table~\ref{tab:audio_params} of the Appendix section. All audio samples were resampled to 22.05 kHz for consistency during training.

To ensure alignment between waveform inputs and spectrogram outputs, we applied a standardized set of audio preprocessing parameters during fine-tuning. These hyperparameters, presented in Table~\ref{tab:audio_params}, define the Short-Time Fourier Transform (STFT) configuration, mel-scale resolution, and audio normalization settings. This configuration is essential to preserve the fidelity of speech characteristics while ensuring compatibility with Tacotron’s temporal and spectral modeling components.

\subsection{Vocoder}

The vocoder component, responsible for converting mel spectrograms into time-domain waveforms, employs a modified WaveRNN architecture \cite{kalchbrenner2018wavernn} comprising an upsampling network, residual convolutional blocks, and recurrent GRU layers optimized for high-fidelity audio synthesis. Due to constraints in computational resources and time, we utilized a pre-trained WaveRNN model and performed fine-tuning on our target dataset rather than training the vocoder from scratch. This approach allowed us to achieve satisfactory synthesis quality while significantly reducing training time and resource requirements.

\subsection{UMAP and EER}

We employ UMAP \cite{mcinnes2018umap} as a powerful nonlinear dimensionality reduction technique to project high-dimensional speaker embeddings into a lower-dimensional space, enabling intuitive visualization and analysis of speaker similarities and clustering patterns. To quantitatively evaluate the biometric verification capability of our speaker encoder, we use the Equal Error Rate (EER) \cite{reynolds2009speaker}, a widely accepted performance metric in speaker recognition that reflects the point at which false acceptance and false rejection rates are equal, providing a balanced measure of system accuracy.

\section{Results}
\label{sec:result}

The performance of the voice cloning system was evaluated across the key components—Speaker Encoder, Synthesizer, and the final perceptual quality assessment using Mean Opinion Score (MOS)\footnote{Sample voices used in this study are available at \url{https://voice-cloning-mos.vercel.app/}}. Each module exhibited stable convergence during training and demonstrated strong generalization and perceptual effectiveness in producing high-fidelity, speaker-consistent synthetic speech.

\subsection{Speaker Encoder}
\begin{figure*}[!t]
    \centering
    \includegraphics[width=0.95\textwidth]{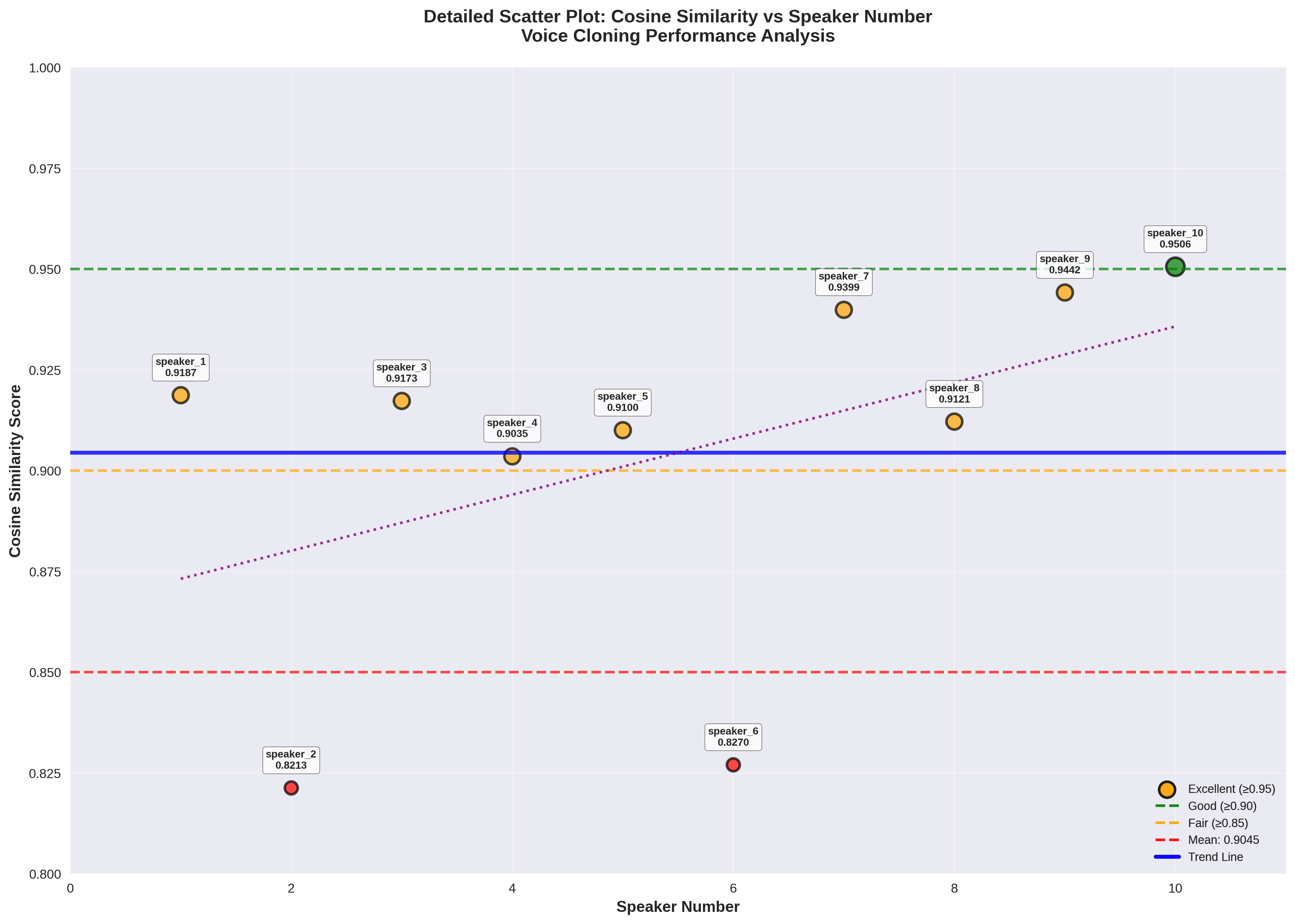}
    \caption{Detailed scatter plot of cosine similarity scores between original and cloned speakers across ten test speakers. The plot highlights performance categories (Excellent, Good, Fair) along with mean and trend lines.}
    \label{fig:cosine_similarity_detailed_scatter}
\end{figure*}

To further assess the quality of the cloned voices, we measured the cosine similarity between the speaker embeddings of the original and cloned samples. A higher cosine similarity indicates closer alignment in the embedding space, and thus higher perceptual similarity between the two voices. Figure~\ref{fig:cosine_similarity_detailed_scatter} presents a detailed scatter plot of cosine similarity scores across ten different speakers.

As shown in Figure~\ref{fig:cosine_similarity_detailed_scatter}, most cloned voices achieved a cosine similarity above 0.90, which is considered a strong indicator of good voice cloning quality. The mean similarity score across all speakers was $0.904457$, marked by the solid blue line. A clear upward trend can be observed (dotted purple line), suggesting consistent improvement as the system adapted to more speakers. 

Performance categorization is also illustrated:  
\begin{itemize}
    \item \textbf{Excellent ($\geq 0.95$):} Speaker\_10 reached the highest similarity ($0.9506$), indicating near-perfect voice cloning.  
    \item \textbf{Good ($\geq 0.90$):} Most speakers, such as Speaker\_1 ($0.9187$), Speaker\_3 ($0.9173$), and Speaker\_9 ($0.9442$), fall in this range.  
    \item \textbf{Fair ($\geq 0.85$):} Speaker\_2 ($0.8213$) and Speaker\_6 ($0.8270$) achieved relatively lower similarity scores, suggesting less accurate cloning performance for these cases.  
\end{itemize}

\begin{figure}[!t]
    \centering
    \includegraphics[width=0.95\columnwidth]{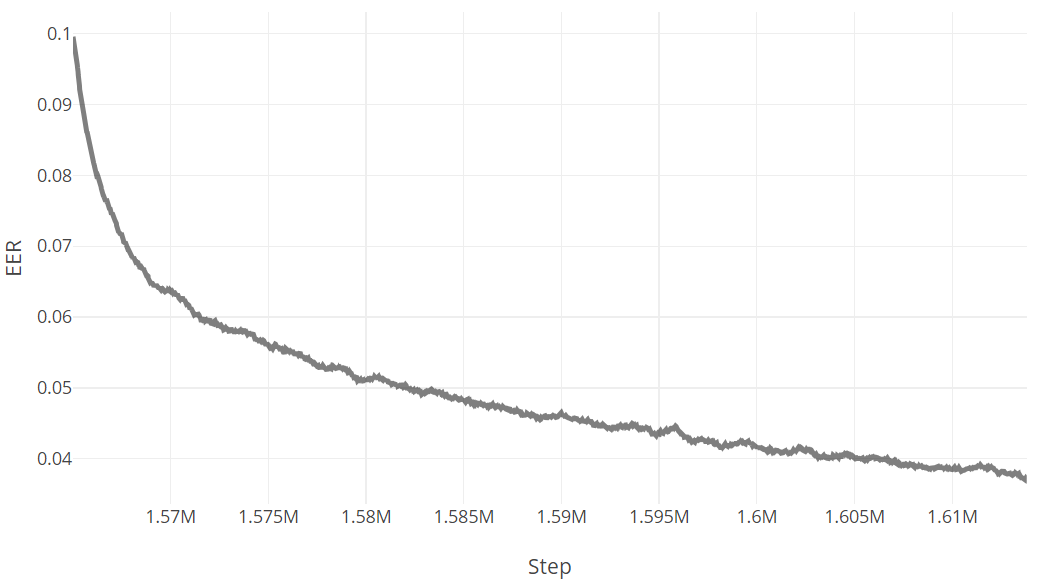}
    \caption{Equal Error Rate (EER) curve during training of the speaker encoder. The steady decline indicates improved discrimination of speaker embeddings over time.}
    \label{fig:eerloss}
\end{figure}

In addition to similarity scores, we evaluated the training progress of the speaker encoder using the Equal Error Rate (EER), a standard metric for speaker verification tasks. As shown in Figure~\ref{fig:eerloss}, the EER steadily decreased with training steps, starting near $0.10$ and dropping below $0.04$. This consistent decline demonstrates that the encoder continually improved its ability to distinguish between speakers, validating the robustness of the learned embeddings.  

\begin{figure}[!t]
    \centering
    \includegraphics[width=0.95\columnwidth]{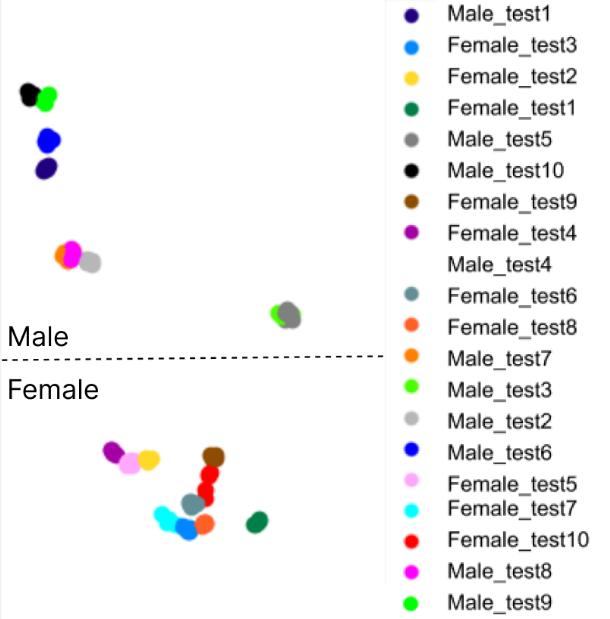}
    \caption{UMAP projection of clustering of different speakers and their embeddings.}
    \label{fig:eer}
\end{figure}

To further examine how well the encoder preserves speaker identity, we visualized the speaker embeddings in a two-dimensional space. Figure~\ref{fig:eer} shows a clustering of embeddings from ten male and ten female test speakers. Each point represents an embedding, and points corresponding to the same speaker are clustered closely together, demonstrating the encoder’s robustness in capturing speaker-specific characteristics. Moreover, a clear separation can be observed between male and female clusters, with male embeddings grouped in the upper region and female embeddings in the lower region. This separation highlights the encoder’s ability to not only cluster voices of the same individual but also capture broader distinctions such as gender, further validating its effectiveness in speaker representation.

\subsection{Synthesizer}

The synthesizer was fine-tuned on a curated dataset of paired text and speech samples. 


\begin{figure*}[t]
    \centering
    \begin{subfigure}[t]{0.45\textwidth}
        \centering
        \includegraphics[width=\textwidth]{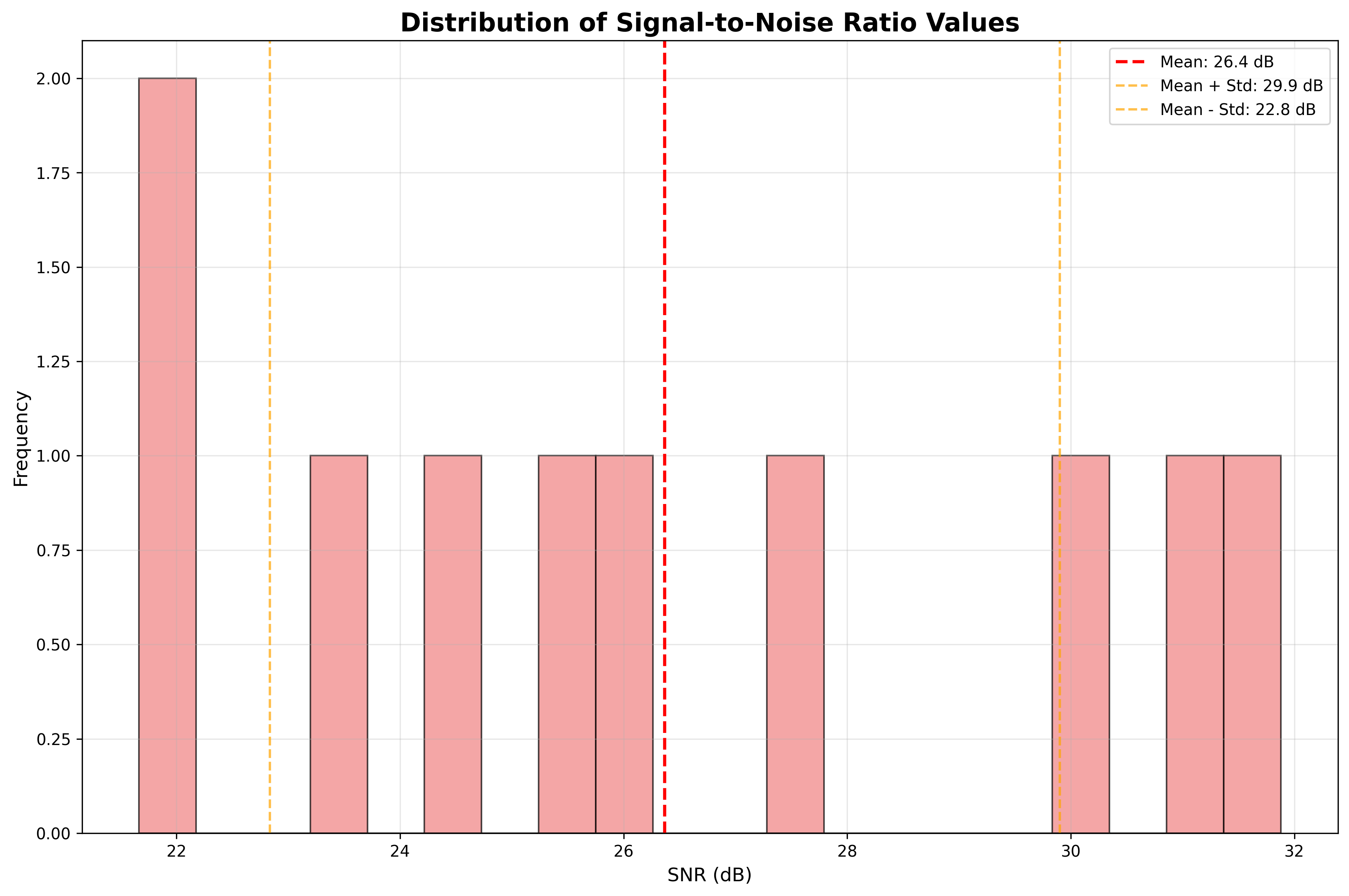}
        \caption{Distribution of Signal-to-Noise Ratio (SNR) values across the dataset. 
        The red dashed line represents the mean SNR (26.4 dB), while the orange dashed 
        lines indicate one standard deviation above and below the mean.}
        \label{fig:snr_histogram}
    \end{subfigure}
    \hfill
    \begin{subfigure}[t]{0.45\textwidth}
        \centering
        \includegraphics[width=\textwidth]{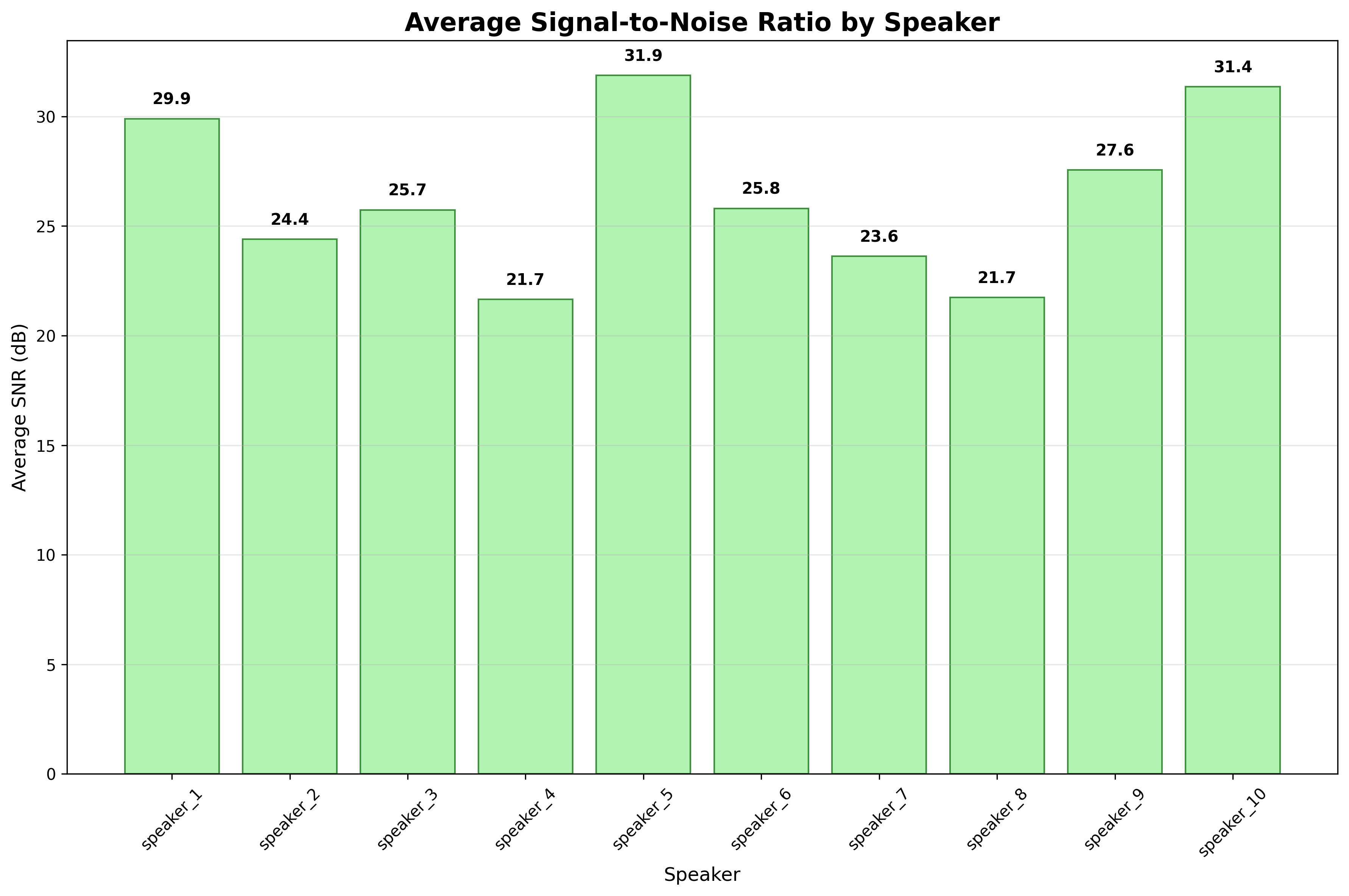}
        \caption{Average Signal-to-Noise Ratio (SNR) for each speaker. The values indicate 
        speaker-specific recording quality, with most speakers exhibiting SNR values above 
        25 dB, and the highest reaching 31.9 dB.}
        \label{fig:snr_speaker}
    \end{subfigure}
    \caption{Signal-to-Noise Ratio (SNR) analysis of the dataset. 
    (a) Distribution of SNR values across the dataset. 
    (b) Average SNR for each speaker.}
    \label{fig:snr_analysis}
\end{figure*}

The quality of the recorded dataset was evaluated using the Signal-to-Noise Ratio (SNR) 
as shown in Figures~\ref{fig:snr_histogram} and~\ref{fig:snr_speaker}. 
Figure~\ref{fig:snr_histogram} illustrates the distribution of SNR values across 
all recordings, where the average SNR is 26.4~dB, with most samples falling 
within one standard deviation (23.6–29.9~dB). This indicates a generally 
consistent recording environment with minimal noise fluctuations. 

Meanwhile, Figure~\ref{fig:snr_speaker} presents the average SNR per speaker. 
The results highlight variation in recording quality across speakers, with SNR values 
ranging from 21.7~dB to 31.9~dB. Notably, most speakers exhibit SNR values above 
25~dB, which demonstrates that the dataset maintains an acceptable quality level for 
robust downstream modeling tasks.

\subsection{Mean Opinion Score Evaluation}

For the perceptual evaluation, we selected 10 speakers—five male and five female—and used three audio clips per speaker with varying durations. A total of 55 participants rated each cloned audio clip based on perceived audio quality and similarity to the corresponding original. Participants were instructed to use headphones for more accurate auditory assessment. The overall MOS ratings averaged $3.924$ for quality and $3.87$ for similarity. As summarized in Table~\ref{tab:mos_summary}, male voices received slightly higher ratings than female ones, though the difference was not substantial. Table~\ref{tab:mos_breakdown} provides a more granular view of scores across individual speakers. Despite variation in speaker characteristics, ratings were relatively consistent, underscoring the system’s robustness. The high standard deviations are attributed to the subjective nature of perceptual evaluation and the relatively small sample of raters.

\begin{table}[!t]
    \centering
    \caption{Summary of the MOS score obtained for our voice cloning}
    \label{tab:mos_summary}
    \small
    \begin{tabular}{lcc}
        \hline
        & \textbf{MOS Quality} & \textbf{MOS Similarity} \\
        \hline
        Male   & 3.84 $\pm$ 0.39 & 3.79 $\pm$ 0.15 \\
        Female & 4.01 $\pm$ 0.34 & 3.95 $\pm$ 0.15 \\
        \hline
    \end{tabular}
\end{table}

A more detailed breakdown of MOS scores by individual speakers is provided in Table~\ref{tab:mos_breakdown}. These results show a high level of consistency in perceived quality and similarity across different speakers, despite some natural variation in vocal characteristics. The relatively large standard deviations observed in the scores can be attributed to the modest participant sample size and the inherent subjectivity of human perception in evaluating synthetic speech.

\begin{table}[!t]
    \centering
    \caption{Breakdown of the MOS score obtained for our voice cloning}
    \label{tab:mos_breakdown}
    \small
    \begin{tabular}{lcc}
        \hline
        \textbf{Speaker} & \textbf{MOS Quality} & \textbf{MOS Similarity} \\
        \hline
        Speaker 1  & 4.01 $\pm$ 0.02 & 3.79 $\pm$ 0.04 \\
        Speaker 2  & 4.10 $\pm$ 0.02 & 3.90 $\pm$ 0.05 \\
        Speaker 3  & 3.57 $\pm$ 0.21 & 3.59 $\pm$ 0.18 \\
        Speaker 4  & 4.22 $\pm$ 0.04 & 4.50 $\pm$ 0.03 \\
        Speaker 5  & 3.60 $\pm$ 0.55 & 3.65 $\pm$ 0.11 \\
        Speaker 6  & 3.83 $\pm$ 0.27 & 4.01$\pm$ 0.16 \\
        Speaker 7  & 3.78 $\pm$ 0.14 & 3.89 $\pm$ 0.18 \\
        Speaker 8  & 4.15 $\pm$ 0.05 & 3.92 $\pm$ 0.07 \\
        Speaker 9  & 3.90 $\pm$ 0.07 & 3.70 $\pm$ 0.07 \\
        Speaker 10 & 3.53 $\pm$ 0.09 & 3.80 $\pm$ 0.16 \\
        \hline
    \end{tabular}
\end{table}

A comprehensive presentation of the experimental results, including additional analyzes and visualizations, is provided in \hyperref[sec:appendixA]{Appendix}. The reader is encouraged to refer to it for detailed metrics, evaluation procedures, and supplementary figures that further illustrate the performance and behavior of the proposed system.

\subsection{Comparative Discussion}
Our system achieves an average MOS of $3.924$ and a speaker similarity score of $3.87$, demonstrating its effectiveness in Nepali multi-speaker voice cloning. A recent study in \cite{karki2024advancingvoicecloningnepali} also explored voice cloning using WaveNet as the vocoder; however, our system outperforms their reported MOS by $0.02$ in quality and $0.67$ in both naturalness and similarity. While their reported scores appear promising, the results are limited to the stated MOS values and calculation procedure, without additional details or supporting evidence provided for verification.
 Furthermore, while their approach relied solely on a pre-existing dataset from the OpenSLR corpus, we constructed a new Nepali text–audio paired corpus, thereby enabling further research opportunities in low-resource language adaptation.  

In comparison, existing Nepali text-to-speech (TTS) systems offer limited functionality. For instance, ``Shruti: A Nepali Book Reader'' \cite{khadka2023tts} focuses primarily on reading digitized book content through OCR-based text extraction, but no perceptual evaluation scores were reported. Similarly, ``Aawaj: Augmentative Communication Support for the Vocally Impaired using Nepali Text-to-Speech'' \cite{aawaj} targets augmentative communication through Nepali TTS, yet it does not support multi-speaker synthesis or provide quantitative performance metrics.  

Compared to these systems, our work broadens the scope of Nepali speech synthesis by introducing multi-speaker voice cloning, while also providing measurable quality and similarity scores. This positions our system as not only functionally distinct but also quantitatively evaluated—an aspect largely absent in previous Nepali TTS and voice cloning research.  

\section{Limitation and Future Enhancements}
\label{sec:limitations}

Our system demonstrates strong performance when the input audio is clear, slightly slower, and features a basic vocal tone, often producing cloned audio that closely resembles the original speaker and allowing listeners to recognize the voice. This highlights the system’s potential; however, several limitations remain. Pronunciation clarity is inconsistent, and the system struggles significantly with unclear audio, noisy environments, or unique linguistic variations, sometimes even generating a completely different or unnatural voice. In such cases, the synthesized speech may exhibit distorted timbre or unnaturally high pitch, reducing naturalness and intelligibility. These shortcomings can be attributed primarily to limited and insufficiently diverse training data, as well as constrained hardware resources. Training was carried out on free or low-resource cloud platforms, which restricted the depth of model optimization and scaling. Furthermore, the dataset—sourced mainly from audiobooks, podcasts, interviews, and OpenSLR, lacked sufficient linguistic diversity, volume, and quality, hindering the system’s ability to generalize across different speech patterns, accents, and vocal characteristics.

For future enhancements, expanding the speaker encoder’s training set is crucial. Our model was trained on fewer than 1,000 speakers, whereas state-of-the-art systems utilize over 9,000 speakers from multiple large-scale corpora, covering diverse native and non-native voices. Increasing both the quantity and diversity of speakers will improve robustness.

Similarly, the synthesizer would benefit from larger, well-transcribed datasets. Our limited collection, constrained by the time-intensive process of aligning audio with transcripts, capped at around 10 hours, restricting TTS quality and naturalness.

Regarding vocoders, while we used WaveRNN due to resource and compatibility constraints, more advanced vocoders like HiFi-GAN or Speaker-Conditional WaveRNN (SC-WaveRNN) offer superior performance, especially for unseen speakers and varied conditions. However, HiFi-GAN demands extensive data and training resources, which were unavailable for this project. SC-WaveRNN, leveraging speaker embeddings, significantly improves audio quality and generalization, and represents a promising direction for future work.

\section{Conclusion}
\label{sec:conclusion}

This research marks an essential first step in extending voice cloning technology to the Nepali language, bridging a critical gap in speech synthesis research. By adapting proven neural architectures to a low-resource setting, we demonstrate the feasibility of developing multi-speaker voice cloning systems for Nepali. Beyond its technical significance, this project underscores the broader cultural value of preserving and amplifying Nepal’s diverse voices by opening avenues for applications in education, accessibility, and digital heritage. While our current system does not yet match the sophistication of those built for high-resource languages, it provides a robust foundation upon which future, more advanced work can be constructed. We hope this endeavor inspires further research that enriches both the technological and cultural landscape of Nepali speech synthesis.

\section{Acknowledgments}
We extend our heartfelt appreciation to the Department of Electronics and Computer Engineering at Thapathali Campus, Institute of Engineering for providing us with the invaluable opportunity to engage in learning.

\bibliographystyle{IEEEtran}
\bibliography{refs}

\clearpage

\appendix
\section*{Appendices}  

\section{Parameters and Hyperparameters Used for Training}
\label{sec:appendixA}

Table~\ref{tab:model_hyperparams} lists the hyperparameters used for training the speaker encoder model. These include the learning rate, embedding and hidden layer sizes, number of LSTM layers, and batch composition in terms of speakers and utterances. Careful tuning of these parameters helps the model effectively learn speaker characteristics from the training data.

\begin{table}[H]
\centering
\caption{Speaker Encoder Model Hyperparameters}
\label{tab:model_hyperparams}
\small
\begin{tabular}{ll}
\hline
\textbf{Parameter} & \textbf{Value} \\
\hline
Initial Learning Rate & $1 \times 10^{-5}$ \\
Embedding Size        & 256 \\
Hidden Layer Size     & 256 \\
Number of LSTM Layers & 3 \\
Speakers per Batch    & 16 \\
Utterances per Speaker & 10 \\
\hline
\end{tabular}
\end{table}

Table \ref{tab:audio_params} summarizes the key audio preprocessing parameters used during synthesizer fine-tuning. These settings, including sampling rate, filter length, and mel filter banks, define how the raw waveform is converted into mel spectrograms, which serve as input features for the model. Proper selection of these parameters ensures accurate frequency representation and temporal resolution, directly impacting the quality and naturalness of the synthesized speech.

\begin{table*}[!t]
\centering
\small
\caption{Audio Preprocessing Parameters Used During Synthesizer Fine-Tuning}
\label{tab:audio_params}
\begin{tabular}{l c p{7.5cm}}
\hline
\textbf{Parameter} & \textbf{Value} & \textbf{Effect on Training} \\
\hline
\texttt{max\_wav\_value} & 32768.0 & Normalizes waveform values from the 16-bit signed integer range \texttt{[-32768, 32767]}, typical for standard PCM audio. \\
\texttt{sampling\_rate} & 22050 & Ensures sufficient temporal resolution to capture all human-audible frequencies without aliasing. \\
\texttt{filter\_length} & 800 & Number of samples used to compute each frame in the STFT. \\
\texttt{hop\_length} & 200 & Sets the step size (in samples) between successive STFT frames, affecting temporal granularity. \\
\texttt{win\_size} & 800 & Size of the window function applied during STFT; defines the analysis frame width. \\
\texttt{n\_mel\_channels} & 80 & Number of mel filter banks used to convert STFT to mel-spectrogram, controlling spectral resolution. \\
\texttt{mel\_fmin} & 0.0 & Lower frequency limit of the mel-spectrogram range. \\
\texttt{mel\_fmax} & 7600 & Upper frequency limit of the mel-spectrogram range. \\
\hline
\end{tabular}
\end{table*}

\section{Additional Results}
\addcontentsline{toc}{subsection}{Additional Results}

\subsection{Male Voice When Text and Transcript Were the Same}

We tested the original and cloned audio using different transcripts—specifically, {\devanagarifont“मैले त आलुको सामान्य परिचय मात्र दिन खोजेको हुँ ।”}, to evaluate the model’s generalization across linguistic variations. This approach assessed not only the clarity and naturalness of the synthesized speech but also the model’s adaptability to changes in pronunciation, intonation, and rhythm across different sentences.

\begin{figure}[!t]
    \centering
    \includegraphics[width=0.95\columnwidth]{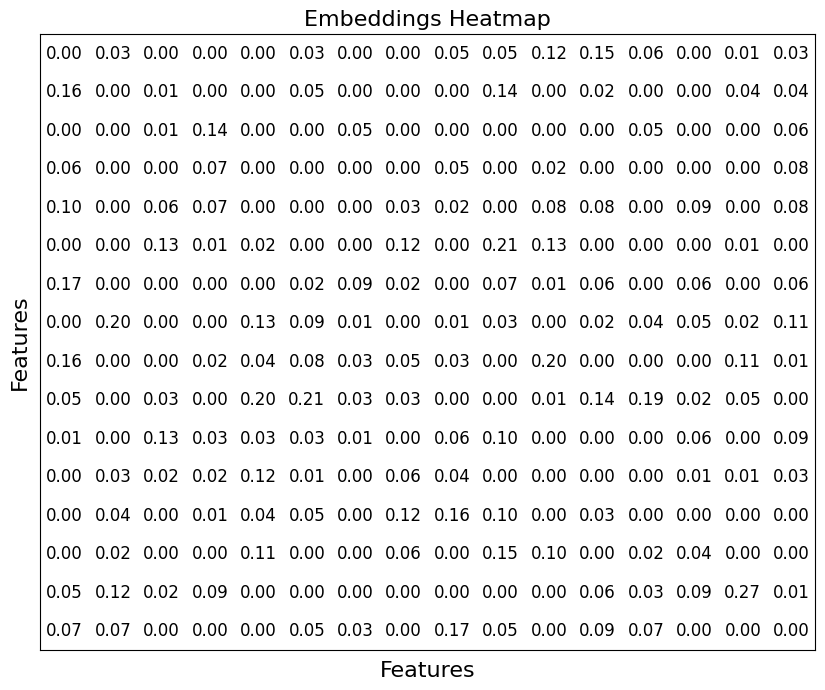}
    \caption{Speaker embedding of Original Male Voice when text and transcript were the same.}
    \label{fig:original-male-embedding}
\end{figure}

\begin{figure}[!t]
    \centering
    \includegraphics[width=0.95\columnwidth]{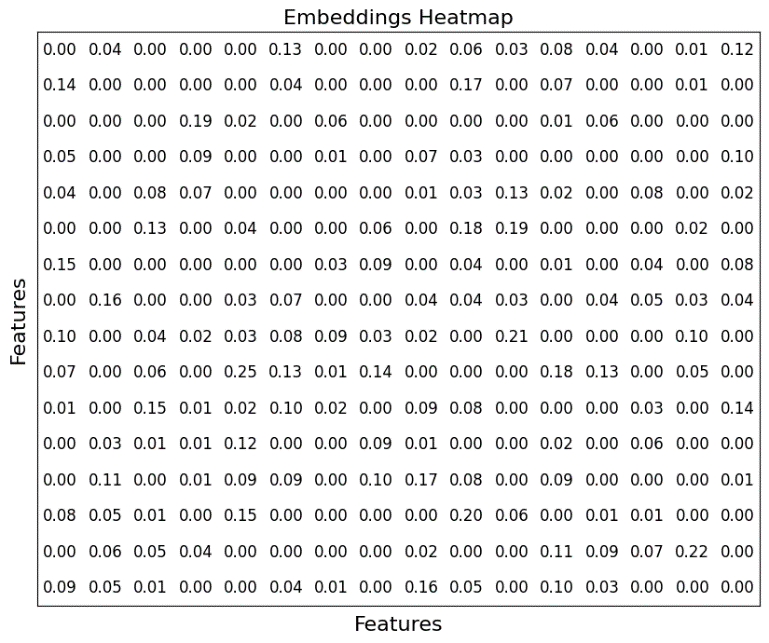}
    \caption{Speaker embedding of Cloned Male Voice when text and transcript were the same.}
    \label{fig:cloned-male-embedding}
\end{figure}

\begin{figure}[!t]
    \centering
    \includegraphics[width=0.95\columnwidth]{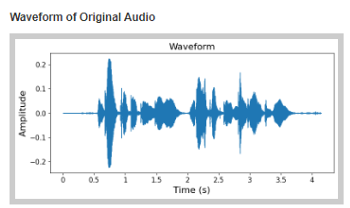}
    \caption{Audio waveform of Original Male Voice when text and transcript were the same.}
    \label{fig:original-male-waveform}
\end{figure}

\begin{figure}[!t]
    \centering
    \includegraphics[width=0.95\columnwidth]{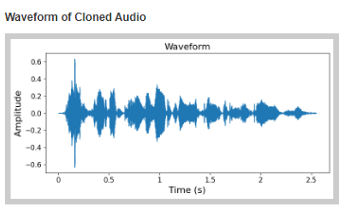}
    \caption{Audio waveform of Cloned Male Voice when text and transcript were the same.}
    \label{fig:cloned-male-waveform}
\end{figure}

The speaker embeddings in Figure \ref{fig:original-male-embedding} and Figure \ref{fig:cloned-male-embedding} show that the cloned audio closely mimics the linguistic features of the original, capturing major speaker characteristics despite minor deviations. While Figure \ref{fig:original-male-waveform} and Figure \ref {fig:cloned-male-waveform}reveals subtle differences in the time-domain waveforms—mainly due to variations in amplitude, phase, and timing—these are expected in speech synthesis and do not significantly affect the perceived quality or similarity, though some minor artifacts are present in the cloned audio.

\begin{figure}[!t]
    \centering
    \includegraphics[width=\columnwidth]{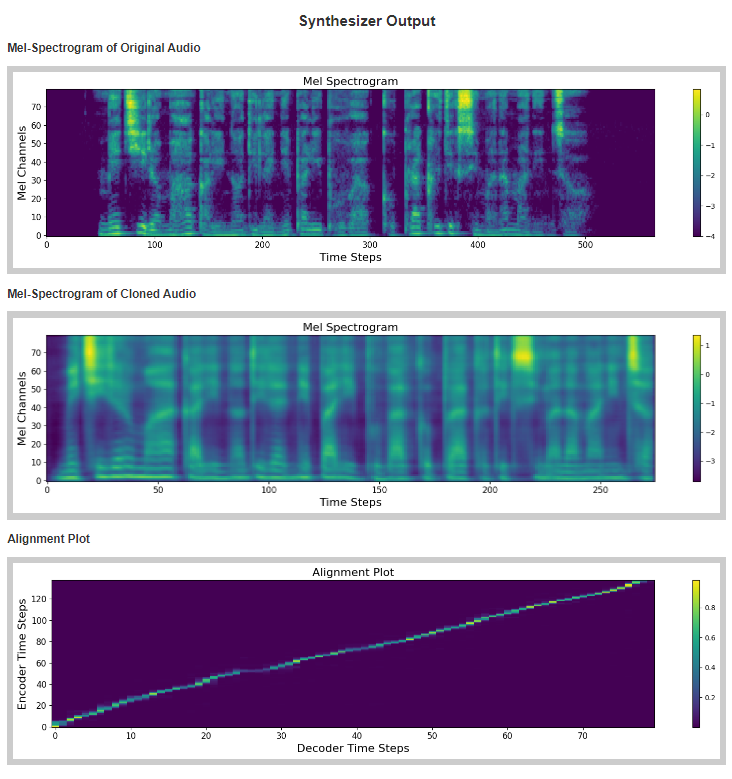}
    \caption{Mel-spectrogram of the original and cloned audio with their alignment plot for male voice when text and transcript were the same.}
    \label{fig:mel-1}
\end{figure}

Figure \ref{fig:mel-1} presents a comparison between the mel spectrograms of the original and cloned audio, along with an alignment plot illustrating synchronization between encoder and decoder time steps. The original audio exhibits a rich frequency range, while the cloned audio retains the general spectral pattern but with reduced sharpness and clarity, indicating slight quality degradation. The alignment plot shows near-linear progression, reflecting good temporal consistency during synthesis

\begin{figure}[!t]
    \centering
    \includegraphics[width=\columnwidth]{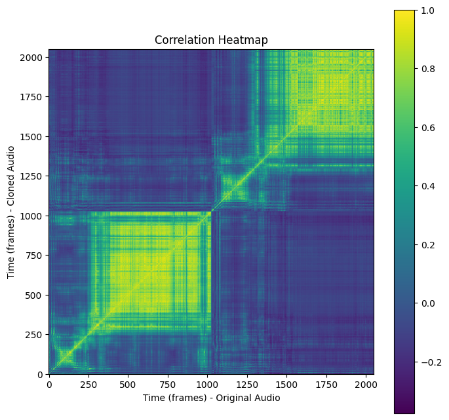}
    \caption{Mel-spectrogram of the original and cloned audio with their alignment plot for male voice when text and transcript were the same.}
    \label{fig:corr-1}
\end{figure}

Figure \ref{fig:corr-1} further supports this with a correlation heatmap, where bright yellow diagonals indicate strong similarity between original and cloned audio segments, and off-diagonal clusters reflect internal similarities with minor temporal shifts, while blue and green areas denote points of dissimilarity.

\subsection{Female Voice When Text and Transcript Were the Same}

We used same transcript for testing both the original and cloned audio, where the text used was {\devanagarifont “उनले कविता र निबन्ध विधामा गरेका योगदानहरू उच्च कोटिको मानिन्छ।”} This allowed us to directly compare the linguistic consistency and acoustic quality between the two audios.

\begin{figure}[!t]
    \centering
    \includegraphics[width=0.95\columnwidth]{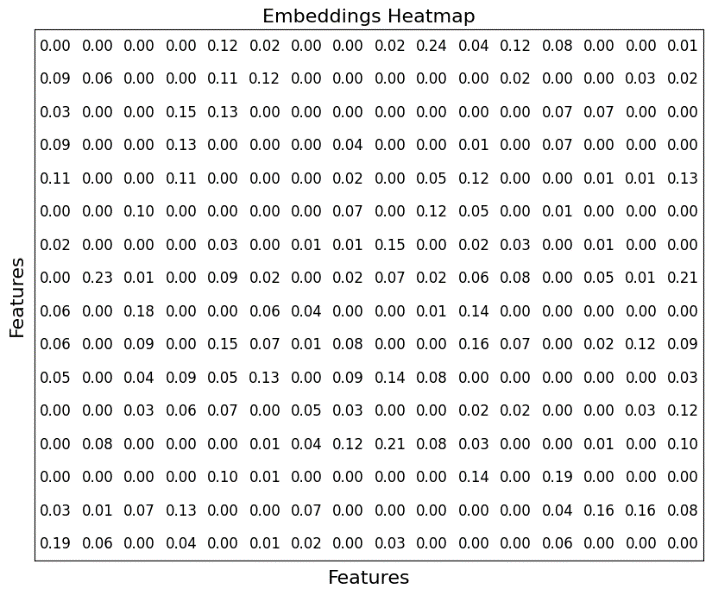}
    \caption{Speaker embedding of Original Female Voice when text and transcript were the same.}
    \label{fig:original-female-embedding}
\end{figure}

\begin{figure}[!t]
    \centering
    \includegraphics[width=0.95\columnwidth]{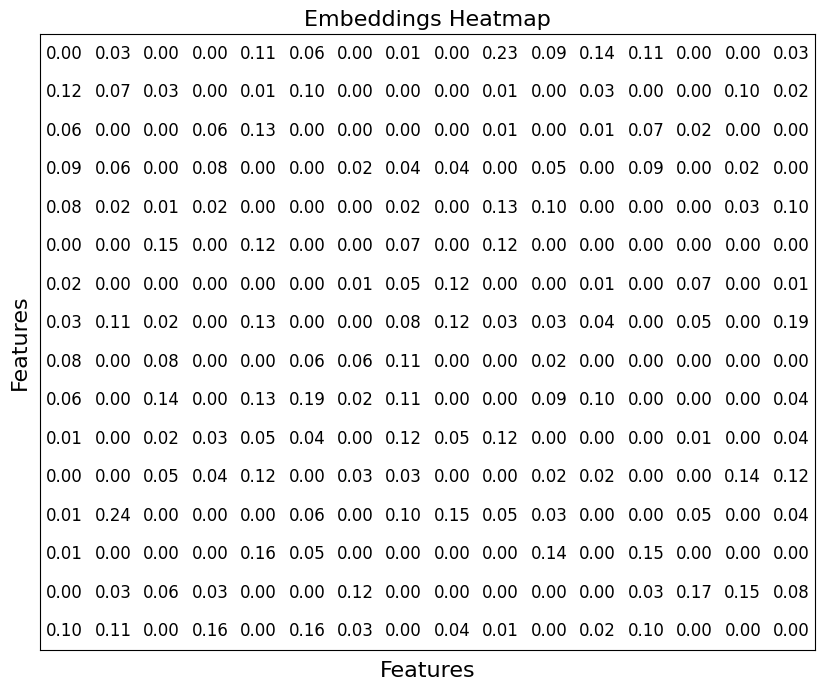}
    \caption{Speaker embedding of Cloned Female Voice when text and transcript were the same.}
    \label{fig:cloned-female-embedding}
\end{figure}

\begin{figure}[!t]
    \centering
    \includegraphics[width=0.95\columnwidth]{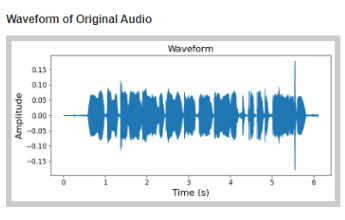}
    \caption{Audio waveform of Original Female Voice when text and transcript were the same.}
    \label{fig:original-female-waveform}
\end{figure}

\begin{figure}[!t]
    \centering
    \includegraphics[width=0.95\columnwidth]{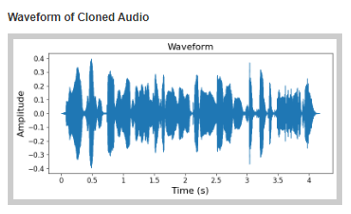}
    \caption{Audio waveform of Cloned Female Voice when text and transcript were the same.}
    \label{fig:cloned-female-waveform}
\end{figure}

The speaker embedding heatmaps in Figure \ref{fig:original-female-embedding} and Figure \ref{fig:cloned-female-embedding} show that while the embeddings aren't identical, they share key features, indicating the model effectively captured and preserved the speaker’s core identity. The cloned audio in Figure \ref{fig:cloned-female-waveform} exhibits slight variations from Figure \ref{fig:original-female-waveform} in duration and waveform spikes, but the overall linguistic similarity remains evident.

\begin{figure}[!t]
    \centering
    \includegraphics[width=\columnwidth]{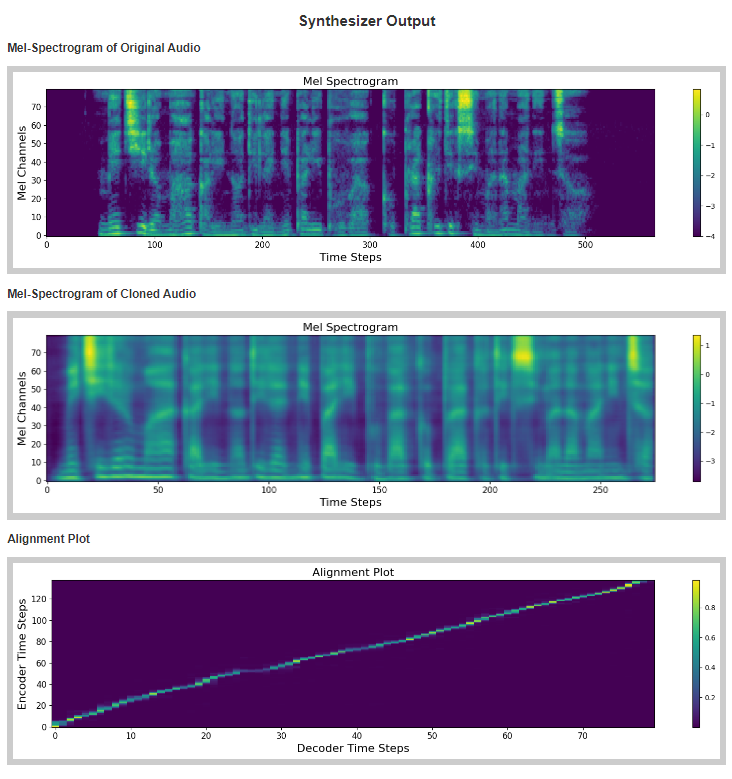}
    \caption{Mel-spectrogram of the original and cloned audio with their alignment plot for female voice when text and transcript were the same.}
    \label{fig:mel-2}
\end{figure}

As shown in Figure \ref{fig:mel-2}, the mel spectrogram and alignment plot of the cloned audio closely resemble those of the original, preserving key spectral features and demonstrating high-fidelity replication despite minor discrepancies. The near-linear alignment plot confirms effective temporal synchronization, ensuring natural rhythm and timing. 

\begin{figure}[!t]
    \centering
    \includegraphics[width=\columnwidth]{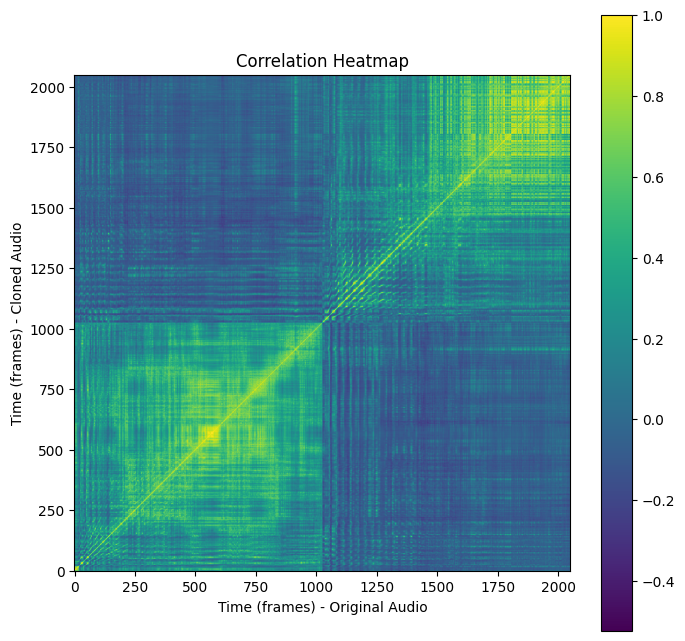}
    \caption{Mel-spectrogram of the original and cloned audio with their alignment plot for female voice when text and transcript were the same.}
    \label{fig:corr-2}
\end{figure}

Figure \ref{fig:corr-2}'s correlation heatmap further supports this, with a strong diagonal indicating high frame-wise similarity, while scattered blue and green regions reflect minor timing mismatches and noise-induced deviations. Overall, the cloned audio replicates the original well with some imperfections.

\subsection{Male Voice When the Text and Transcript Differed}

We used different transcript for testing both the original and cloned audio, where the text used was {\devanagarifont “मैले त आलुको सामान्य परिचय मात्र दिन खोजेको हुँ।“} 

\begin{figure}[!t]
    \centering
    \includegraphics[width=0.95\columnwidth]{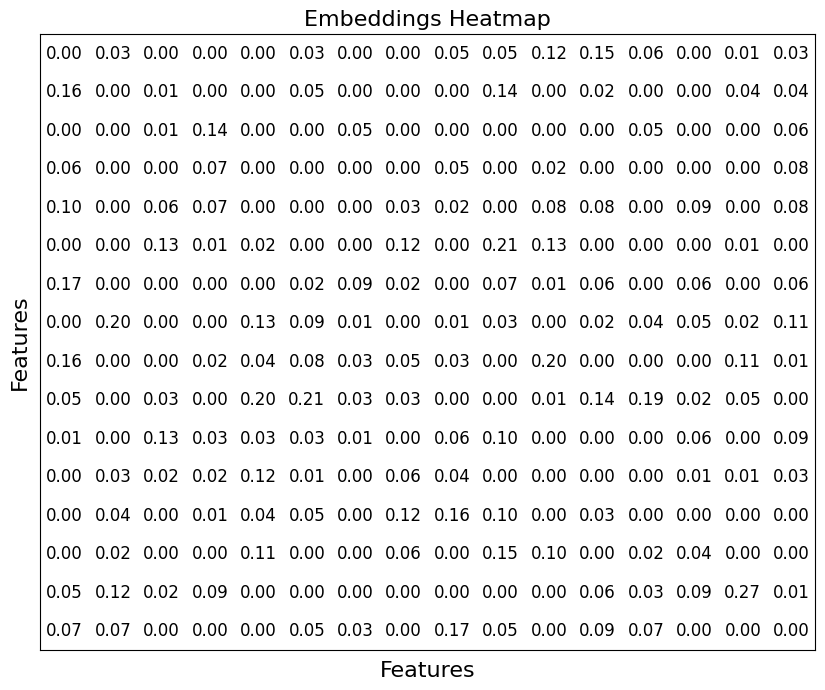}
    \caption{Speaker embedding of Original Male Voice when text and transcript differed.}
    \label{fig:original-male-embedding-diff}
\end{figure}

\begin{figure}[!t]
    \centering
    \includegraphics[width=0.95\columnwidth]{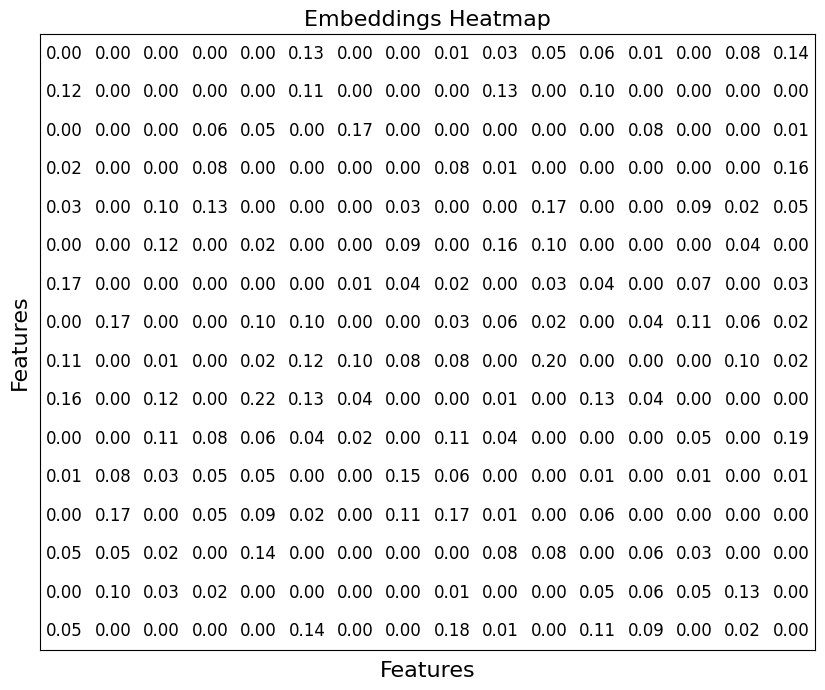}
    \caption{Speaker embedding of Cloned Male Voice when text and transcript differed.}
    \label{fig:cloned-male-embedding-diff}
\end{figure}

\begin{figure}[!t]
    \centering
    \includegraphics[width=0.95\columnwidth]{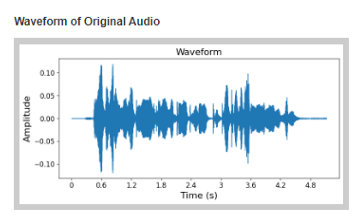}
    \caption{Audio waveform of Original Male Voice when text and transcript differed.}
    \label{fig:original-male-waveform-diff}
\end{figure}

\begin{figure}[!t]
    \centering
    \includegraphics[width=0.95\columnwidth]{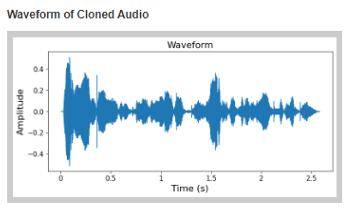}
    \caption{Audio waveform of Cloned Male Voice when text and transcript differed.}
    \label{fig:cloned-male-waveform-diff}
\end{figure}

Figures \ref{fig:original-male-embedding-diff} and Figure \ref{fig:cloned-male-embedding-diff} representing embeddings and Figure \ref{fig:original-male-waveform-diff} and Figure \ref{fig:cloned-male-waveform-diff} representing waveform, show that despite different texts and transcripts, the speaker embeddings reveal notable similarities between the original and cloned speakers. This indicates that the cloned audio captures the main characteristics of the original speaker, even when generated from different text. The time-domain waveform shows no distortion in the cloned audio; differences in amplitude reflect the text variation but do not introduce significant noise or artifacts, demonstrating successful cloning.

\begin{figure}[!t]
    \centering
    \includegraphics[width=\columnwidth]{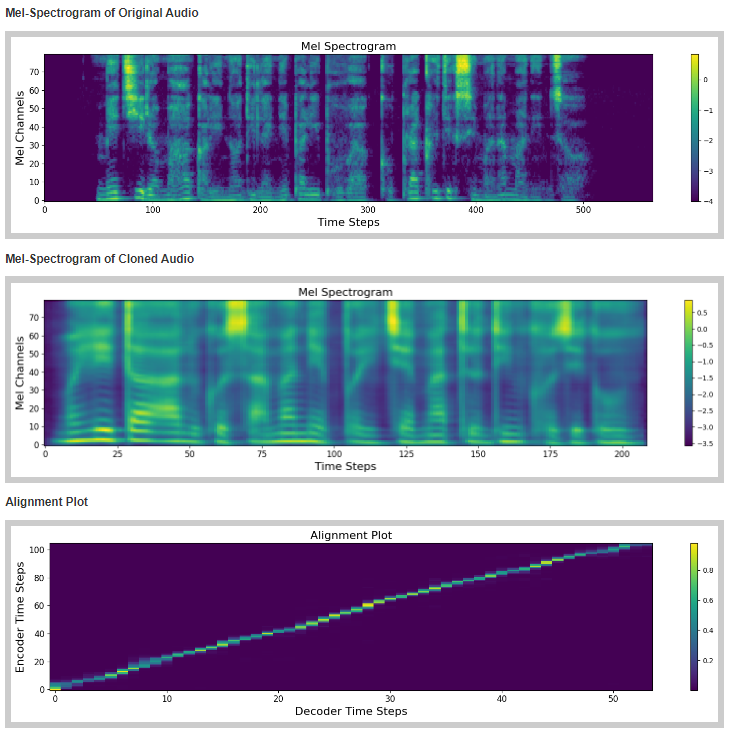}
    \caption{Mel-spectrogram of the original and cloned audio with their alignment plot for male voice when text and transcript differed.}
    \label{fig:mel-3}
\end{figure}

The mel spectrogram of the original audio, as shown in Figure \ref{fig:mel-3}, displays a rich and dynamic frequency range, capturing various elements such as vowels and consonants with differing intensity levels that reflect the speaker’s vocal quality. Although the cloned audio’s mel spectrogram differs due to the use of different text, it still effectively preserves the major vocal characteristics of the original speaker. The alignment plot exhibits a diagonal pattern, indicating a linear relationship between encoder and decoder time steps, which demonstrates that the model maps input to output in a coherent and structured way.

\subsection{Female Voice When the Text and Transcript Differed}

We used the different transcript for testing both the original and cloned audio, where the text used was {\devanagarifont “मेची र महाकाली नदीलाई नेपाली भूभागको प्राकृतिक सिमाना मानिन्छ।”} This allowed us to directly compare the linguistic consistency and acoustic quality between the two audios.

\begin{figure}[!t]
    \centering
    \includegraphics[width=0.95\columnwidth]{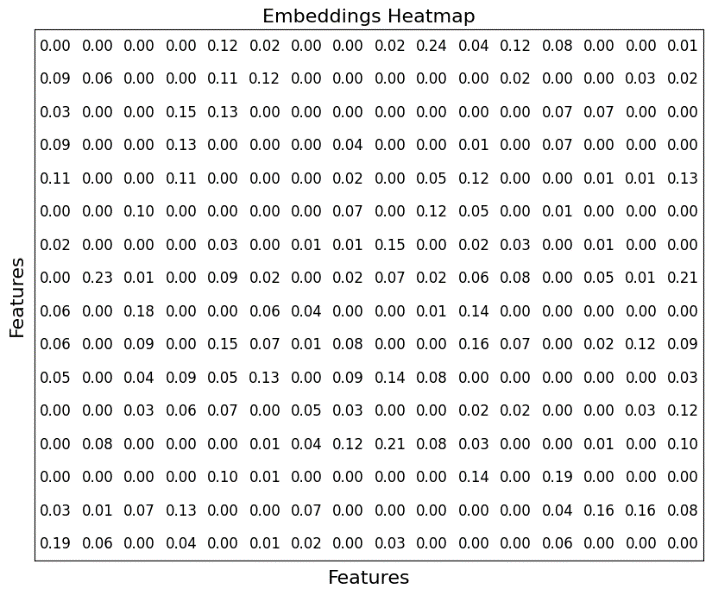}
    \caption{Speaker embedding of Original Female Voice when text and transcript differed.}
    \label{fig:original-female-embedding-diff}
\end{figure}

\begin{figure}[!t]
    \centering
    \includegraphics[width=0.95\columnwidth]{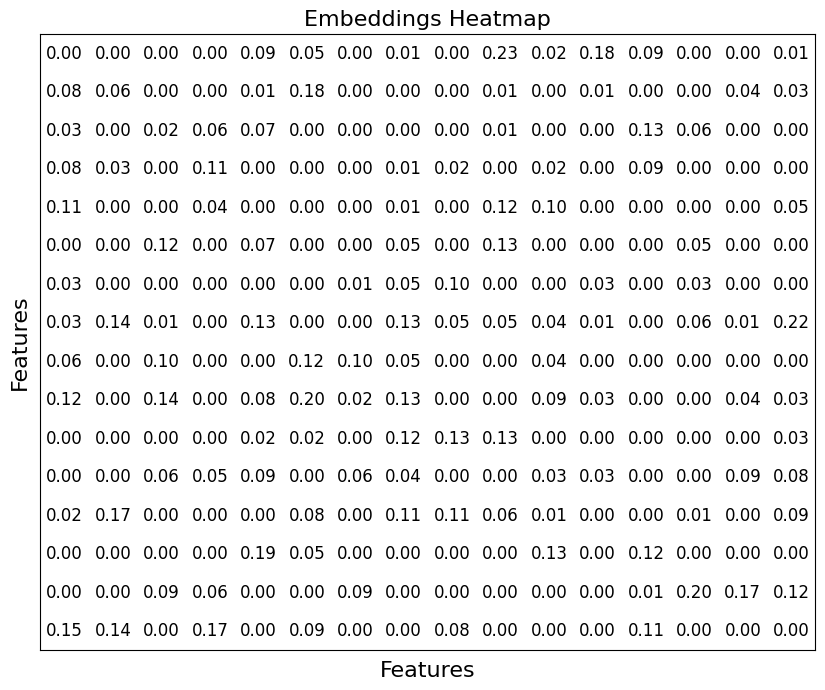}
    \caption{Speaker embedding of Cloned Female Voice when text and transcript differed.}
    \label{fig:cloned-female-embedding-diff}
\end{figure}

\begin{figure}[!t]
    \centering
    \includegraphics[width=0.95\columnwidth]{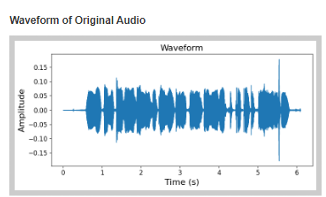}
    \caption{Audio waveform of Original Female Voice when text and transcript differed.}
    \label{fig:original-female-waveform-diff}
\end{figure}

\begin{figure}[!t]
    \centering
    \includegraphics[width=0.95\columnwidth]{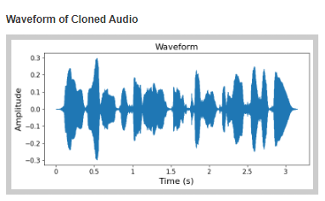}
    \caption{Audio waveform of Cloned Female Voice when text and transcript differed.}
    \label{fig:cloned-female-waveform-diff}
\end{figure}

They had very similar linguistic features for medium text as well, as shown in the embedding graphs in Figure \ref{fig:original-female-embedding-diff} and Figure \ref{fig:cloned-female-embedding-diff}. Even when the transcripts for the original and cloned audio differed, the speaker embeddings did not vary significantly. This similarity allows the cloned audio to capture major features like pitch and rhythm of the original speaker and reproduce them effectively.

The time-domain waveform in Figure \ref{fig:cloned-female-waveform-diff} confirms that no distortion occurred during the generation of the cloned audio. Similar to the male waveform, differences arise from the differing texts used, with some spikes reflecting amplitude variations, but no significant noise is present in the cloned audio.

\begin{figure}[!t]
    \centering
    \includegraphics[width=\columnwidth]{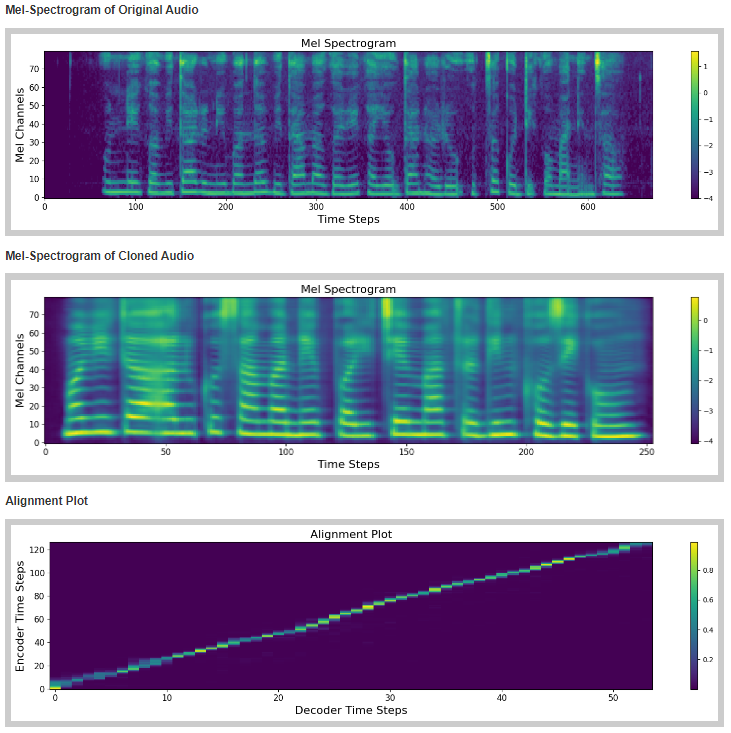}
    \caption{Mel-spectrogram of the original and cloned audio with their alignment plot for female voice when text and transcript differed.}
    \label{fig:mel-4}
\end{figure}

The cloned audio showed no distortion or disturbance, as evident from the mel spectrograms and alignment plot in Figure \ref{fig:mel-4}. Both the original and cloned audio cover a similar frequency range, visible in the top and middle spectrograms. The alignment plot demonstrates that the encoder and decoder time steps maintain a nonlinear relationship, supporting the natural timing of the cloned audio.

\end{document}